\documentclass[11pt]{article}
\usepackage{amsmath,epsfig,a4,cite,latexsym,color,axodraw
}
\usepackage{mathrsfs}

\newcommand{\fxx}{w}

\newcommand{\fLiComb}{f_{\text{H}}}
\newcommand{\amub}{a_{\mu;\text{bos}}^{\rm EW(2)}}
\newcommand{\amuf}{a_{\mu;\text{ferm}}^{\rm EW(2)}}

\newcommand{\amufrestH}{a_{\mu;\text{f-rest,H}}^{\rm EW(2)}}
\newcommand{\amufrestHgamma}{a_{\mu;\text{f-rest,H}\gamma}^{\rm EW(2)}}
\newcommand{\amufrestHZ}{a_{\mu;\text{f-rest,HZ}}^{\rm EW(2)}}
\newcommand{\amufrestnoH}{a_{\mu;\text{f-rest,no H}}^{\rm EW(2)}}

\def\pslash#1{{\setbox0=\hbox{$#1$}
    \rlap{\ifdim\wd0>.7em\kern.22\wd0\else\kern.1\wd0\fi /}#1}}

\begin{document}
\begin{flushright}
\end{flushright}
\vspace{3em}
\begin{center}
{\Large\bf The electroweak 
contributions to 
$(g-2)_\mu$
after the  \\[1ex] Higgs boson mass measurement}
\\
\vspace{3em}

{\sc C.\ Gnendiger,  D.\ St\"ockinger,
H.\ St\"ockinger-Kim
}\\[2em]
{\sl Institut f\"ur Kern- und Teilchenphysik,
TU Dresden, Dresden, Germany}
\setcounter{footnote}{0}
\end{center}
\vspace{2ex}
\begin{abstract}
The Higgs boson mass used to be the only unknown input parameter of
the electroweak contributions to $(g-2)_\mu$ in the Standard Model.
It enters at the two-loop level in diagrams with e.g.\ top loops, W-
or Z-exchange. We re-evaluate these 
contributions, providing analytic expressions and exact numerical
results for the Higgs boson mass recently measured at the LHC.
Our final result for the full Standard Model
electroweak contributions is $(153.6\pm1.0)\times10^{-11}$, where the
remaining theory error comes from unknown three-loop contributions and
hadronic uncertainties. 
\end{abstract}

\vspace{0.5cm}

The anomalous magnetic moment $a_\mu=(g-2)_\mu/2$ of the muon has been
measured very precisely at Brookhaven National Laboratory, with the final
value\cite{BNL2006}\footnote{
The change in the number compared to Ref. \cite{BNL2006} is due to
a new PDG value for the magnetic moment ratio of the muon to proton,
see e.g.\ Ref.\ \cite{FNALProposal}}:
\begin{align}
a_\mu^{\rm exp} &= (116\,592\,089 \pm 63)\times 10^{-11} \, .
\label{amuexp}
\end{align}
This measurement has already reached a sensitivity to
details of the weak interactions, which contribute at the order $10^{-9}$.
Future experiments planned at Fermilab \cite{FNALProposal} and J-PARC \cite{Iinuma:2011zz} aim
to further reduce the uncertainty by a factor~4.

The Standard Model theory prediction has also been continuously
improving, see Refs.\ \cite{JegerlehnerNyffeler,MdRRS} for recent
reviews and references.
The 5-loop QED contribution has been completely calculated
\cite{Kinoshita2012}. The hadronic vacuum polarization contributions 
make use of the most recent experimental data on the ($e^+e^-\to\ $hadrons)
cross section \cite{Davier,HMNT,Benayoun:2012wc}, and an earlier discrepancy to analyses
based on $\tau$-decays has been resolved \cite{JegerlehnerSzafron,Benayoun:2012wc}.
The latest results of various groups for the hadronic light-by-light contributions agree
within the quoted errors \cite{JegerlehnerNyffeler,dRPV}, and new non-perturbative approaches
promise further progress \cite{Goecke:2010if,Blum:2013qu}.

Here we focus on the electroweak contributions to $(g-2)_\mu$ in the Standard
Model. They include contributions from the Higgs boson and are the
only ones which depend on the Higgs boson mass $M_H$. This quantity
used to be the only unknown input parameter of the Standard Model,
resulting in the dominant remaining theory uncertainty of the
electroweak contributions.
As a reference, the seminal evaluation of Ref.\
\cite{CzMV} obtained the result
\begin{align}
a_\mu^{\rm EW}&=(154\pm1\pm2)\times10^{-11},
\end{align}
where the first error is due to hadronic uncertainties, but the second
is due to the unknown Higgs boson mass.

Now, the Higgs boson mass has been measured at the LHC to be
$M_H =125.5 \pm 0.2 ({\rm stat.}) {}^{+0.5}_{-0.6} ({\rm syst.})\ {\rm GeV}$ by ATLAS \cite{ATLAS:2013mma} and
$M_H =125.7 \pm 0.3 ({\rm stat.}) \pm 0.3 ({\rm syst.})\ {\rm GeV}$ by CMS\cite{CMS:yva}.
In the following we take the average
central value and a conservative error band, covering the $2\sigma$
range of both measurements:
\begin{align}
M_H=125.6\pm1.5\ {\rm GeV}.
\label{MHinput}
\end{align}

Given this progress on all fronts regarding $(g-2)_\mu$ and the Higgs boson
it is appropriate to update the prediction of the electroweak
contributions to $(g-2)_\mu$.

In the present paper we therefore re-evaluate the electroweak Standard
Model contributions at the two-loop level, making use of the LHC result.
We provide the full
$M_H$-dependent part in numerical and, where not readily available, in
analytical form. This allows us to obtain the exact $(g-2)_\mu$ prediction for the
measured value of $M_H$, and to compare with previously published results
and error estimates. We combine this with the most advanced
computations of all other electroweak contributions up to leading
3-loop order and provide the final result and a complete discussion of the
remaining theory error.

In the following our input parameters besides Eq.\
(\ref{MHinput}) are \cite{PDG2012}:
\begin{subequations}
\begin{align}
m_\mu &= 105.6583715 \pm 0.0000035\ \mbox{MeV},\\
M_Z &= 91.1876 \pm 0.0021\ \mbox{GeV}, \\
m_t &= 173.5\pm 0.6 \pm 0.8\ \mbox{GeV}
\end{align}
\end{subequations}
for the masses of muon, Z-boson and top quark, and
\begin{subequations}
\begin{align}
G_F &= (1.166\, 378\, 7\pm 0.000\, 000\, 6) \times 10^{-5}\ \mbox{GeV}^{-2},\\
\alpha &=1/137.035\, 999&
\label{GFAlphaDef}
\end{align}
\end{subequations}
for the muon decay constant and the fine-structure constant.
Given these parameters the W-boson mass is predicted by the
Standard Model theory \cite{Awramik:2003rn}. We obtain%
\footnote{
In \mbox{Ref.\ \cite{Ferroglia:2012ir}} another top quark mass
has been used: $m_t=173.2\pm 0.9$ GeV. There $M_W$ equals 
$80.361\pm 0.010$ GeV.
}
\begin{align}
M_W &= 80.363 \pm 0.013\ \mbox{GeV}.
\label{Wmass}
\end{align}
\\

The Standard Model
electroweak contributions are split up into one-loop, two-loop
and higher orders as 
\begin{align}
\label{EWsplitup}
a_\mu^{\rm EW} &=
a_\mu^{\rm EW(1)} + \amub + \amuf + a_\mu^{\rm EW(\ge3)},
\end{align}
where the two-loop contributions are further split into bosonic and
fermionic contributions, as discussed below.

The one-loop contribution is given by \cite{JegerlehnerNyffeler,MdRRS}%
\footnote{
In the literature sometimes the experimental value for $M_W$ instead of
the theory value is used. If we would use the current value
of $M_W=80.385\pm0.015$ GeV \cite{PDG2012} instead of
\mbox{Eq.\ \eqref{Wmass}}, the result would be shifted to
$a_\mu^{\rm EW(1)}=(194.81\pm 0.01)\times 10^{-11}$.
}
\begin{align}
a_\mu^{\rm EW(1)}&=\frac{G_F}{\sqrt 2} \frac{m_\mu^2}{8\pi^2}\left[\frac53 +
\frac13(1-4s_W^2)^2\right]=(194.80\pm 0.01)\times 10^{-11},
\label{oneloop}
\end{align}
where $s_W^2=1-M_W^2/M_Z^2$ is
the square of the weak mixing angle in the on-shell renormalization
scheme. One-loop contributions suppressed by $m_\mu^2/M_Z^2$ or
$m_\mu^2/M_H^2$ are smaller than $10^{-13}$ and hence neglected here.
The parametrization in terms of $G_F$ already absorbs important
higher-order contributions. The error in Eq.\ (\ref{oneloop}) is due
to the uncertainty of the input parameters, in particular of the W-boson
mass.

Before discussing higher-order contributions we briefly explain
possible parametrizations in terms of $G_F$ and $\alpha$. 
The one-loop contribution in \mbox{Eq.\ (\ref{oneloop})} has been parametrized
in terms of $G_F$.
Generally, $n$-loop contributions are proportional to $G_F\,\alpha^{(n-1)}$, and it
is possible to reparametrize $\alpha$ in terms of other quantities.
Possibilities are to replace $\alpha$ by a running $\alpha$ at the
scale of the muon mass or the Z-boson mass, or to replace $\alpha\to\alpha(G_F)$,
where $\alpha(G_F)\equiv \sqrt{2}G_F s_W^2 M_W^2/\pi=\alpha\times(1+\Delta r)$.
The quantity $\Delta r$ summarizes radiative corrections to muon decay.
Different choices amount to differences which are formally of the order $n+1$.
We will always choose $\alpha$ in the Thomson limit, i.e.\ given by Eq.\ (\ref{GFAlphaDef}).
\begin{figure}
\begin{center}
\begin{tabular}{cccc}

\scalebox{0.55}{
\begin{picture}(150,90)(0,0)
\ArrowLine(0,0)(25,0)
\Vertex(25,0){2}
\DashLine(25,0)(45,30){3}
\Photon(45,30)(65,60){2}{3.5}
\Photon(65,60)(105,0){2}{6.5}
\Photon(45,30)(65,0){2}{4}
\ArrowLine(25,0)(65,0)
\ArrowLine(65,0)(105,0)
\Vertex(105,0){2}
\ArrowLine(105,0)(130,0)
\Text(10,-10)[]{\Large$\mu$}
\Text(120,-10)[]{\Large$\mu$}
\Text(25,20)[]{\Large$H$}
\Text(100,30)[]{\Large$W$}
\Text(0,60)[]{\huge(a)}
\Photon(65,60)(65,90){2}{3.5}
\Text(75,80)[]{\Large$\gamma$}
\end{picture}
}


&
\scalebox{0.55}{
\begin{picture}(150,90)(0,0)
\ArrowLine(0,0)(40,0)
\Vertex(40,0){2}
\DashLine(40,0)(40,30){3}
\ArrowLine(40,0)(90,0)
\Vertex(90,0){2}
\Photon(90,0)(90,30){2}{3.5}
\ArrowLine(90,0)(130,0)
\Text(10,-10)[]{\Large$\mu$}
\Text(120,-10)[]{\Large$\mu$}
\Text(30,15)[]{\Large$H$}
\Text(110,15)[]{\Large$ \gamma$, $Z$}
\Text(0,60)[]{\huge(b)}

\ArrowLine(90,30)(40,30)
\ArrowLine(40,30)(65,60)
\ArrowLine(65,60)(90,30)
\Text(65,40)[]{\Large$f$}

\Photon(65,60)(65,90){2}{3.5}
\Text(75,80)[]{\Large$\gamma$}
\end{picture}
}

&
%
%
\scalebox{0.55}{
\begin{picture}(150,90)(0,0)
\ArrowLine(0,0)(40,0)
\Vertex(40,0){2}
\Photon(40,0)(40,30){3}{3.5}
\ArrowLine(40,0)(90,0)
\Vertex(90,0){2}
\Photon(90,0)(90,30){2}{3.5}
\ArrowLine(90,0)(130,0)
\Text(10,-10)[]{\Large$\mu$}
\Text(120,-10)[]{\Large$\mu$}
\Text(30,15)[]{\Large$Z$}
\Text(100,15)[]{\Large$\gamma$}
\Text(0,60)[]{\huge(c)}
\ArrowLine(90,30)(40,30)
\ArrowLine(40,30)(65,60)
\ArrowLine(65,60)(90,30)
\Text(65,40)[]{$f$}
\Photon(65,60)(65,90){2}{3.5}
\Text(75,80)[]{\Large$\gamma$}
\end{picture}
}
&
\scalebox{0.55}{
\begin{picture}(150,90)(0,0)
\ArrowLine(0,0)(25,0)
\Vertex(25,0){2}
\ArrowLine(25,0)(65,60)
\ArrowLine(65,60)(105,0)
\ArrowArc(65,0)(15,0,180)
\ArrowArc(65,0)(15,180,360)
\Photon(25,0)(50,0){2}{2.5}
\Photon(80,0)(105,0){2}{2.5}
\Vertex(105,0){2}
\ArrowLine(105,0)(130,0)
\Text(10,-10)[]{\Large$\mu$}
\Text(120,-10)[]{\Large$\mu$}
\Text(35,30)[]{\Large$\mu$}
\Text(95,30)[]{\Large$\mu$}
\Text(37,-10)[]{\Large$Z$}
\Text(93,-10)[]{\Large$\gamma$}
\Text(65,0)[]{\Large$f$}
\Text(0,60)[]{\huge(d)}
\Photon(65,60)(65,90){2}{3.5}
\Text(75,80)[]{\Large$\gamma$}
\end{picture}
}
\end{tabular}
\end{center}
\caption{\label{fig:diagrams} Sample two-loop diagrams: Higgs-dependent bosonic (a) and
 fermionic (b) diagram, diagram with $\gamma\gamma Z$-fermion triangle (c)
 and $\gamma$--$Z$ mixing (d).}
\end{figure}
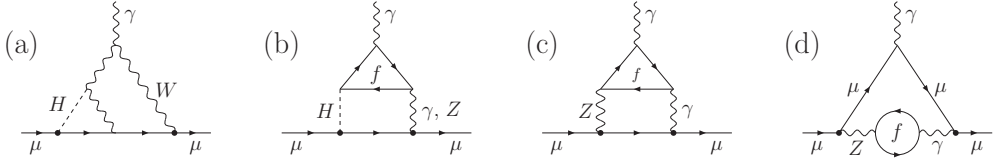

We now turn to the first set of contributions with noticeable
dependence on the Higgs boson mass: the bosonic two-loop contributions
$\amub$. They are defined by two-loop and
associated counterterm diagrams without a closed fermion
loop, see  Fig.~\ref{fig:diagrams}(a) for a sample diagram.
They are conceptually straightforward but involve many diagrams. 
Their first full computation in Ref.\ \cite{CKM2} was a milestone ---
the first full computation of a Standard Model observable at the
two-loop level. Actually, Ref.\ 
\cite{CKM2} employed an approximation assuming $M_H\gg M_W$. Ref.\
\cite{HSW04} confirmed the result but provided the full
$M_H$-dependence; Ref.\
\cite{CzarneckiGribouk} then published the result in semianalytical
form. 
\begin{figure}
\includegraphics[width=0.5\textwidth]{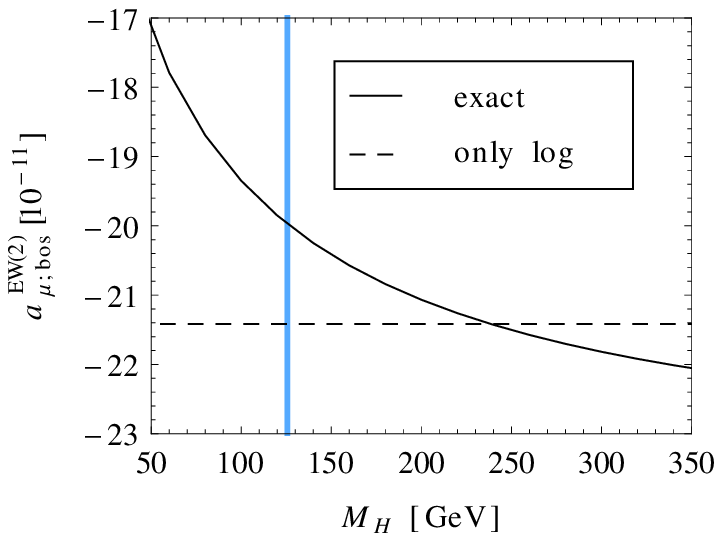}
\includegraphics[width=0.5\textwidth]{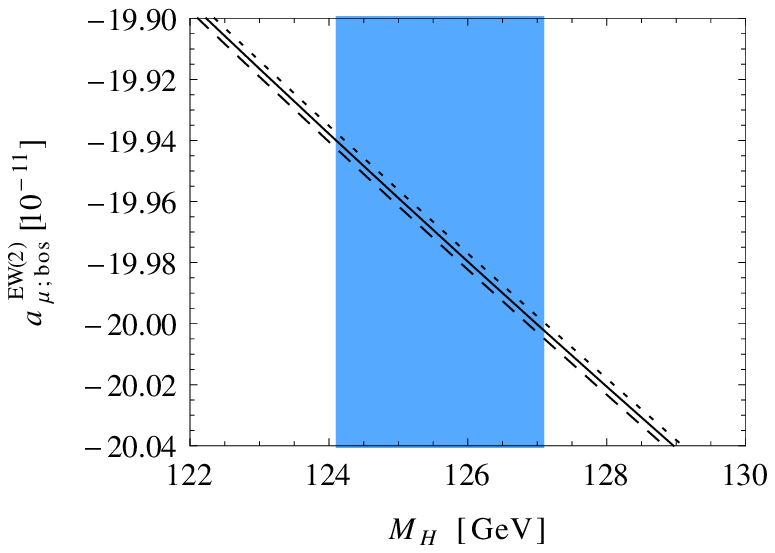}
\caption{Numerical result for $\amub$ 
  as a function of the Higgs boson mass. The vertical
  band indicates the measured value of $M_H$. The dashed line
  in the left plot corresponds to the leading logarithmic
  approximation as defined in Ref.\ \cite{HSW04}. In the right plot
  the dotted, solid, dashed lines correspond to a variation of $M_W$
  by $(-15,0,+15)$ MeV, respectively.
}
\label{fig:Higgsplotsbos}
\end{figure}

Here we re-evaluate the bosonic two-loop contributions using the
parametrization discussed above, in terms of $G_F\,\alpha$. 
Fig.\ \ref{fig:Higgsplotsbos} shows
the result for a range of Higgs boson masses. The numerical result
differs by around $3\%$ from the one given in 
Ref.\ \cite{HSW04}, where the $G_F\,\alpha(G_F)$ parametrization
was chosen. 
The measured value of $M_H$ now fixes the value of these contributions
and we obtain 
\begin{align}
\label{Higgsbosonicres}
\amub&=(-19.97\pm 0.03)\times 10^{-11}.
\end{align}
Here the remaining parametric uncertainty results from the experimental
uncertainties of the input parameters $M_H$, and to a smaller extent of 
$M_W$, see the right plot in Fig.\ \ref{fig:Higgsplotsbos}.
The result lies within the intervals given in the original
Refs.\ \cite{HSW04,CzarneckiGribouk} and the recent reviews
\cite{JegerlehnerNyffeler,MdRRS}, which all differ slightly because of
the different Higgs boson mass ranges and central values used for the
evaluations.

The fermionic two-loop contributions $\amuf$
are defined by Feynman diagrams with a closed fermion loop. 
The Higgs boson enters through diagrams of the type of
Fig.~\ref{fig:diagrams}(b), where a fermion loop generates a
$H\gamma\gamma$ or $H\gamma Z$ interaction.
The fermionic contributions involve also light quark loops, e.g.\
in the diagrams of
Fig.~\ref{fig:diagrams}(c), for which
perturbation theory is questionable. Hence we split up these contributions
further, slightly extending the notation of Ref.\ \cite{MdRRS}:
\begin{align}
\amuf &= 
a_\mu^{\rm EW(2)}(e,\mu,u,c,d,s)
+a_\mu^{\rm EW(2)}(\tau,t,b)
+\amufrestH
 + \amufrestnoH \, .
\end{align}

Here the first two terms on the r.h.s.~denote contributions from the
diagrams of Fig.\ \ref{fig:diagrams}(c) with a
$\gamma\gamma Z$-subdiagram and the indicated fermions in the loop.
The third term denotes the Higgs-dependent diagrams of
Fig.~\ref{fig:diagrams}(b); the fourth collects all remaining
fermionic contributions, e.g.\ from W-boson exchange or from diagram
Fig.~\ref{fig:diagrams}(d).

We first focus on the Higgs-dependent part, for which we write 
\begin{align}
\amufrestH =
\sum_{f} \left[\amufrestHgamma(f)+\amufrestHZ(f)\right] ,
\end{align}
where the two terms in the sum denote the Higgs-dependent diagrams of
Fig.~\ref{fig:diagrams}(b) with either a photon or a Z-boson in the
outer loop and the sum extends over the Standard Model fermions; the
relevant ones are
$f=t,b,c,\tau$. Contributions from the remaining Standard Model
fermions are below $10^{-14}$ and thus negligible.

The first full computation of the fermionic contributions, including the Higgs
dependence was carried out in Ref.\ \cite{CKM1}. There, the dependence
on the Higgs boson 
mass is provided in three limiting cases, \mbox{$M_H\ll m_t$},
\mbox{$M_H= m_t$}, \mbox{$M_H\gg m_t$}. Furthermore, since
$s_W^2\approx1/4$, terms suppressed by a factor \mbox{$(1-4s_W^2)$},
in particular the entire Higgs--Z diagrams of Fig.\ \ref{fig:diagrams}(b) were
neglected.
Diagrams similar to Fig.~\ref{fig:diagrams}(b) have also been
evaluated in the more complicated case of extended models, e.g.\ in
the Two-Higgs-doublet model and the supersymmetric Standard Model
\cite{Cheung:2001hz,Wu:2001vq}.  

We computed the Higgs-dependent diagrams without approximations in two
ways: with the technique developed for Ref.\ \cite{HSW03,HSW04}
using asymptotic expansion and integral reduction techniques, and with
the method of Barr and Zee, where the inner loop is computed
first and then inserted into the outer loop \cite{BarrZee}.
The result from this is
\begin{align}
\label{beginfermrestHresults}
\amufrestHgamma(f)
=&\,
\frac{G_{F}}{\sqrt 2}\frac{m_{\mu}^{2}}{8 \pi ^2}\frac{\alpha}{\pi}\,
N_{C}\,Q_{f}^{2}
\,\, 2 \, f_{H\gamma}(x_{fH}),\\
\amufrestHZ(f)
=&\, 
\frac{G_{F}}{{\sqrt 2}} \frac{m_{\mu}^{2}}{8 \pi ^{2}} \frac{\alpha}{\pi} \, N_{C}\,Q_{f}
\frac{I_{f}^{3} -2 s_{W}^2 Q_{f}}{4 c_{W}^{2} s_{W}^{2}}(1 - 4 s_{W}^2)
\, \, f_{HZ}(x_{fH},x_{fZ}),
\end{align}
with $x_{fH} = m_{f}^{2}/M_{H}^{2}$ and $x_{fZ}=m_{f}^2/M_Z^2$. The
loop functions can be written in terms of one-dimensional integral
representations or in terms of dilogarithms:
\begin{align}
f_{H\gamma}(x) =& 
\int_{0}^{1} d\fxx \, x\, \frac{2 \fxx^{2} - 2 \fxx + 1}{\fxx^2 - \fxx +
  x} \log \frac{\fxx ( 1 - \fxx)}{x} 
\\
=&\,x\, \left[\fLiComb(x)-4\right],
\label{fHgamma}\\
f_{HZ}(x,z)=&
\int _{0}^{1} d\fxx \, x\,z\,\frac{2 \fxx^2 - 2 \fxx + 1}{\fxx^2 - \fxx +z}
\left[ \frac{\log\frac{\fxx(1-\fxx)}{x}}{\fxx^{2} - \fxx + x} +
\frac{\log\frac{x}{z}}{x-z} \right]\\
=& \frac{x\,z}{x-z} \left[\fLiComb(z)-\fLiComb(x)
\label{fHZ}
\right].
\end{align}
The dilogarithms are contained in the function $\fLiComb(x)$, defined
as\footnote{
  In Ref.\ \cite{Stockinger:2006zn}, Eq.(70), a similar
  function $f_{S}(x)$ is defined, where \mbox{$f_{S}(x) = x \fLiComb(x)-4x$}.
  Additionally, Eqs.\ (\ref{fHgamma}), (\ref{fHZ}) are connected by $f_{H\gamma}(x) = \lim_{z\rightarrow\infty}f_{HZ}(x,z)$.
  }
\begin{align}
\fLiComb(x)=\frac{4x-2}{y}
\left[
\text{Li}_2\left(1-\frac{1-y}{2x}\right)-
\text{Li}_2\left(1-\frac{1+y}{2x}\right)
\right]-2\log x,
\end{align}
with $y=\sqrt{1-4x}$.
Further, the weak isospin $I^3_f$ is defined as $\pm\frac12$ for up (down)
fermions, and the electric charge $Q_{f}$ equals $+\frac23,-\frac13,-1$ for
up-type quarks, down-type quarks and charged leptons, respectively.
The color factor $N_{C}$ is $1$ for leptons and $3$ for quarks.
\begin{figure}
\includegraphics[width=0.5\textwidth]{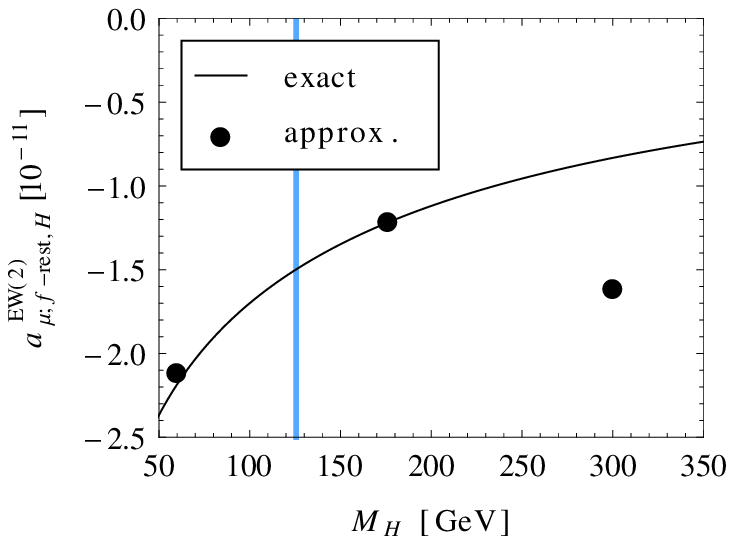}
\includegraphics[width=0.5\textwidth]{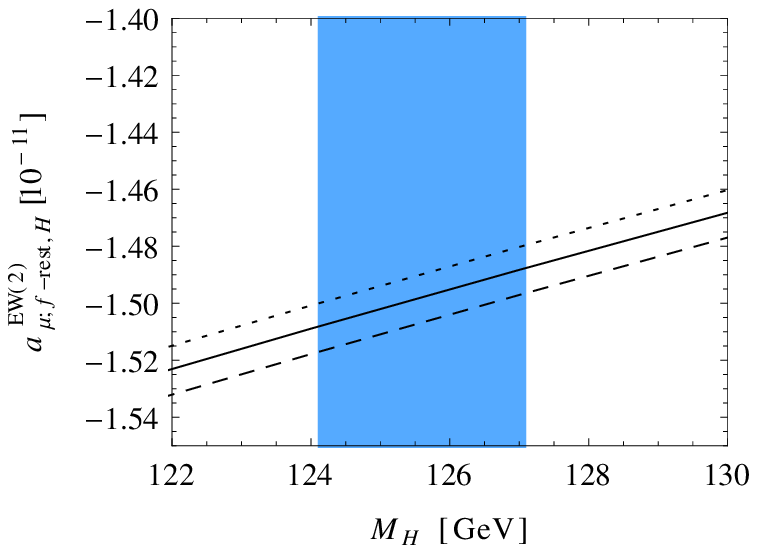}
\caption{Numerical result for $\amufrestH$
  as a function of the Higgs boson mass. The vertical
  band indicates the measured value of $M_H$. The fat dots
  in the left plot correspond to the approximations for
  $M_H=60\ \mbox{GeV},m_t,300\ \mbox{GeV}$ given in Ref.\ \cite{CKM1}. In
  the right plot 
  the dotted, solid, dashed lines correspond to a variation of $m_t$
  by $(-1.4,0,+1.4)$ GeV, respectively.}
\label{fig:Higgsplotsferm} 
\end{figure}

Fig.\ \ref{fig:Higgsplotsferm}(a) shows the numerical result as a function of the
Higgs boson mass and compares with the numerical values obtained in
Ref.\ \cite{CKM1}, using their approximations.
We find that the approximation for large $M_H$ is surprisingly poor. 
As a check of this case, we have explicitly computed the higher orders in the expansion in $m_{t}^{2}/M_{H}^{2}$ and 
verified that the terms neglected in Ref.\ \cite{CKM1} are important. 

Inserting the measured value of the Higgs boson mass, and taking into
account all contributions including top, bottom, charm and $\tau$
loops and diagrams with Higgs and Z-boson exchange, we obtain
\begin{align}
\amufrestH=(-1.50\pm0.01)\times10^{-11},
\label{Higgsfermionicres}
\end{align}
where the indicated error arises essentially from the uncertainty of
the input parameters $m_t$ and $M_H$.
Again, the result is in agreement with the intervals given in Refs.\
\cite{CKM1,JegerlehnerNyffeler,MdRRS}, which differ because of the
different allowed Higgs boson mass ranges.

Eqs.\ (\ref{Higgsbosonicres}), (\ref{beginfermrestHresults})--(\ref{Higgsfermionicres}) and Figs.\
\ref{fig:Higgsplotsbos} and \ref{fig:Higgsplotsferm} constitute our
main new results. In the following we briefly review the remaining
electroweak contributions, with slight updates.

The non-Higgs dependent contributions $\amufrestnoH$ are given by:
\begin{align}
\begin{split}
\amufrestnoH
=&- \frac{G_F}{\sqrt{2}}\frac{m_\mu^2}{8\pi^2}\frac{\alpha}{\pi}
\bigg[
\frac{1}{2s_W^2}\left(\frac{5}{8}\frac{m_t^2}{M_W^2}+
\log\frac{m_t^2}{M_W^2}+
\frac{7}{3}\right)\bigg] \\
&-\frac{G_F}{\sqrt{2}}\frac{m_\mu^2}{8\pi^2}\frac{\alpha}{\pi}
\bigg[
\frac{c_W^2}{2s_W^2}\frac{m_t^2}{M_W^2}
\left(1-4s_W^2\right)\bigg] \\
&-\frac{G_F}{\sqrt{2}}\frac{m_\mu^2}{8\pi^2}\frac{\alpha}{\pi}
\bigg[\bigg.
\left(\frac{8}{9}\log\frac{M_Z}{m_\mu}+\frac{4}{9}\log\frac{M_Z}{m_\tau}\right)
\left(1-4s_W^2\right)^2 \\ 
&\quad\quad\quad\quad\quad\quad+
\frac{4}{3}\times 6.88
\left(1-4s_W^2\right)
\bigg.\bigg].
\end{split}
\label{fermrestnoHiggsanalytical}
\end{align}

The first line has been computed in Ref.\ \cite{CKM1} and was re-written in this form e.g.\ in 
Ref.\ \cite{JegerlehnerNyffeler,Knechtreview}; the other two terms correspond to additional terms
added in Ref.\ \cite{CzMV}, where however no explicit formula was provided.
These terms are suppressed by $(1-4s_W^2)$ but enhanced by either $m_t^2/M_W^2$
or by large logarithms. 
The factor $m_t^2/M_W^2$ enters via the
quantity $\Delta \rho$, which arises by applying the
renormalization $s_W^2\to s_W^2+\delta s_W^2$ in the
$(1-4s_W^2)^2$-term of the one-loop result~(\ref{oneloop}).
The other term originates from diagrams with
$\gamma$--$Z$ mixing as shown in Fig.~\ref{fig:diagrams}(d)
with light fermions running in the loop.
It can be computed using renormalization-group techniques \cite{DG98,CzMV}.
The number $6.88$ in the last line has been obtained in Ref.\ \cite{CzMV} as a
nonperturbative replacement of the perturbative expression
$2/3\sum_{q=u,d,s,c,b} N_c\left(I^3_q Q_q-2Q_q^2 s_W^2\right)\text{log} M_Z/m_q$.
Numerically, we obtain $-4.12, -0.23, -0.29$ in units of $10^{-11}$ for
the three contributions, in total 
\begin{align}
\amufrestnoH=(-4.64\pm0.10)\times10^{-11}.
\label{fermrestnoHiggs}
\end{align}
The error due to the uncertainty of the input parameters 
is negligible; the given error is our estimate of the still neglected
terms which are suppressed by a factor $(1-4s_W^2)$ or $M_Z^2/m_t^2$ and not
enhanced by anything. The estimate is obtained by comparison with the
computed terms in the second and third line of Eq.\
(\ref{fermrestnoHiggsanalytical}) and 
the respective enhancement factors. 

\begin{figure}
\centering
\includegraphics[width=0.5\textwidth]{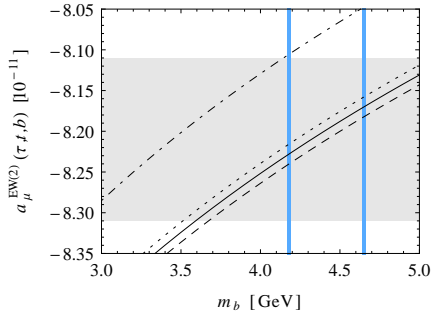}
\caption{Numerical result for $a_\mu^{\rm
    EW(2)}(\tau,t,b)$ as a function of the bottom quark mass, for
  various values of the top quark mass. The dash-dotted line
  corresponds to the $\overline{\mbox{MS}}$ mass $\overline{m}_t=160$ GeV;
    the solid, dotted, dashed lines to the pole mass $m_t=173.5$ GeV
    and variations thereof by $\mp1.4$ GeV. The vertical
  blue lines indicate the values of $\overline{\mbox{MS}}$ mass
  $\overline{m}_b(\overline{m}_b)=4.18$ GeV and the
  1S-mass $4.65$ GeV \cite{PDG2012}; the $\overline{\mbox{MS}}$ mass at higher scales
  has smaller values. The horizontal gray band corresponds to the
  result (\ref{EWVVA3}).}
\label{fig:amuferm3rd}
\end{figure}

For the third generation contributions to
Fig.\ \ref{fig:diagrams}(c) perturbation theory can be
applied, and these contributions have been evaluated in Refs.\
\cite{Peris:1995bb,CKM1,CzMV}. The result and the error
estimate from Ref.\ \cite{CzMV}, including subleading 
terms in $m_t^2/M_Z^2$, read
\begin{align}
\label{EWVVA3}
a_\mu^{\rm EW(2)}(\tau,t,b)&=-(8.21\pm0.10)\times10^{-11}.
\end{align}
We have re-evaluated these
contributions for various definitions of quark
masses which differ by higher orders in the strong interaction,
similarly to the error estimation by Ref.\ \cite{CzMV}. The result is
shown in Fig.\ \ref{fig:amuferm3rd}, and it confirms that Eq.\
(\ref{EWVVA3}) is still compatible with present values of quark
masses.

The contribution of the first two generations to Fig.\
\ref{fig:diagrams}(c) has first been fully computed in Ref.\ \cite{CKM1},
approximating the light quark contributions by a naive perturbative
calculation with constituent-like quark masses.
The treatment of the light quark contributions has been successively improved
in later references, by taking into account non-perturbative information
on the longitudinal \cite{Peris:1995bb,Knecht:2002hr}, then on both the
longitudinal and transverse parts 
of the $\gamma\gamma Z$ three-point function \cite{CzMV}. The final
result of Ref.\
\cite{CzMV} is%
\footnote{The result is taken
  from the erratum of Ref.\ \cite{CzMV}. It is
  perfectly compatible with the one provided in Ref.\
  \cite{JegerlehnerNyffeler}. The result quoted in Ref.\ \cite{MdRRS}  was taken from
  the original Ref.\ \cite{CzMV}; it differs slightly but  is also
  compatible within the errors.}
\begin{align}
\label{EWVVA12}
a_\mu^{\rm EW(2)}(e,\mu,u,c,d,s)=-(6.91\pm0.20\pm0.30)\times10^{-11},
\end{align}
where the uncertainties for the 1st and 2nd generation have been given
separately.

Contributions from beyond the two-loop level have been considered in
Refs.\ \cite{DG98,CzMV}. There, the leading
logarithms at the three-loop level have been obtained from
renormalization-group methods. It was found that these logarithms
amount to $0.4\times10^{-11}$, if the two-loop result is parametrized
in terms of $G_F\,\alpha(m_\mu)$, where $\alpha(m_\mu)$ is the
running fine-structure constant at the scale of the muon mass. If the
two-loop result is parametrized in terms of $G_F\,\alpha$, however, the
shift of the coupling accidentally cancels the three-loop logarithms. 
Hence, since this is the parametrization we have used, we take
\begin{align}
a_\mu^{\rm EW(\ge3)}=(0\pm0.20)\times10^{-11},
\label{EWthreeloop}
\end{align}
where the error estimate is from Ref.\
\cite{CzMV}. It corresponds to estimating the
non-leading logarithmic three-loop contributions to be below a percent
of the two-loop contributions.\\

\begin{figure}
\centering
\includegraphics[width=0.5\textwidth]{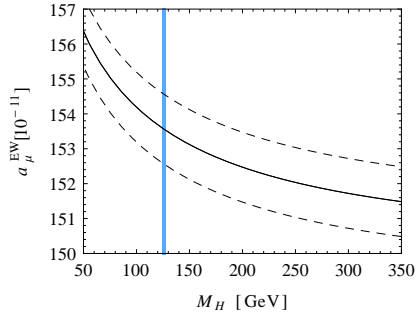}
\caption{
Numerical result for $a_\mu^{\rm EW}$ as a function of the Higgs boson mass.
The vertical band indicates the measured value of $M_H$.
The dashed lines correspond to the uncertainty of the final result, 
quoted in \mbox{Eq.\ \eqref{amuEWNew}}.}
\label{fig:amuEWfull}
\end{figure}

In summary, we have re-evaluated the electroweak contributions to 
$a_\mu$ using the measured Higgs boson mass and
employing consistently the $G_F\,\alpha$ parametrization at the
two-loop level. We provide exact numerical results for the full
bosonic and the Higgs-dependent fermionic two-loop contributions, for
the latter also analytical results. These results are supplemented by
updates of the most advanced available results on all other electroweak
contributions. Our final result obtained from
Eqs.\
(\ref{oneloop}),\ (\ref{Higgsbosonicres}),\ %
(\ref{Higgsfermionicres}),\ (\ref{fermrestnoHiggs}),\ %
(\ref{EWVVA3}),\ (\ref{EWVVA12}),\ (\ref{EWthreeloop})
reads
\begin{align}
a_\mu^{\rm EW}&= (153.6\pm1.0)\times10^{-11}
\label{amuEWNew}
\end{align}
and is illustrated in \mbox{Fig.\ \ref{fig:amuEWfull}}.
We assess the final theory error of these contributions to be
$\pm1.0\times10^{-11}$. This is the same value as the one
given in Ref.\ \cite{CzMV} for the overall hadronic uncertainty from the
diagrams of Fig.\ \ref{fig:diagrams}(c), which is 
now by far the dominant source of error of the electroweak contributions.
The error from unknown three-loop contributions and neglected two-loop terms
suppressed by $M_Z^2/m_t^2$ and $(1-4s_W^2)$ is significantly smaller and the error due to the
experimental uncertainty of the Higgs boson, W-boson, and top-quark
mass is well below $10^{-12}$ and thus negligible.

%
Our result is consistent with the previous evaluations of the
electroweak contributions in Refs.\
\cite{JegerlehnerNyffeler,MdRRS,CzMV}, whose central values range
between  \mbox{$(153\ldots 154)\times10^{-11}$}, but the large uncertainty
due to the unknown Higgs boson mass has been reduced.
In comparison, the recent 5-loop calculation \cite{Kinoshita2012}
has shifted the QED result by $+0.8\times10^{-11}$.
We can now combine Eq.\ \eqref{amuEWNew} and the result of 
Ref.\ \cite{Kinoshita2012} with the 
hadronic contributions. We take the recent leading order evaluations of 
Refs.\ \cite{Davier} and \cite{HMNT} and the higher order results of
Refs.\ \cite{HMNT,dRPV}.
The resulting difference between the experimental result
Eq.\ (\ref{amuexp}) and the full Standard Model prediction is:
\begin{align}
a_\mu^{\rm exp}-a_\mu^{\rm SM} &=
\begin{cases}
(287\pm80)\times10^{-11} \ \mbox{\cite{Davier}}, \\
(261\pm80)\times10^{-11} \ \mbox{\cite{HMNT}}.
\end{cases}
\end{align}

The Standard Model theory error remains dominated by the
non-electroweak hadronic contributions. 
The QED and electroweak contributions can now be regarded as
sufficiently accurate for the precision of next generation 
experiments.

\subsection*{Acknowledgements}
Communications with A. Czarnecki, E. de Rafael and B. Lee Roberts are
gratefully acknowledged.
This work has been supported by the German Research Foundation DFG through
Grant No. STO876/1-1.




\begin{thebibliography}{AA}
\bibitem{BNL2006}G.W. Bennett, et al.,
(Muon $(g-2)$ Collaboration), Phys. Rev. D {\bf 73}, 072003 (2006).
%
%
\bibitem{FNALProposal}D.~W.~Hertzog, B.~Lee
Roberts et al., Fermilab Proposal P-989, March 2009,
{\tt{\verb+http://www.fnal.gov/directorate/program_planning/+ \verb+Mar2009PACPublic/PACMarch09AgendaPublic.htm+}};
  B.~L.~Roberts,
  arXiv:1001.2898 [hep-ex].
%
\bibitem{Iinuma:2011zz}
  H.~Iinuma [J-PARC New g-2/EDM experiment Collaboration],
  J.\ Phys.\ Conf.\ Ser.\  {\bf 295} (2011) 012032.
%
\bibitem{JegerlehnerNyffeler}
  F.~Jegerlehner and A.~Nyffeler,
  Phys.\ Rept.\  {\bf 477} (2009) 1.
%
\bibitem{MdRRS} J.\ Miller, E.\ de Rafael, B.L.\ Roberts, D.\
  St\"ockinger, Ann.Rev.Nucl.Part. (2012) 62.
%
\bibitem{Kinoshita2012}
  T.~Aoyama, M.~Hayakawa, T.~Kinoshita and M.~Nio,
  Phys.\ Rev.\ Lett.\  {\bf 109} (2012) 111808
  [arXiv:1205.5370 [hep-ph]].
%
\bibitem{Davier}
  M.~Davier, A.~Hoecker, B.~Malaescu and Z.~Zhang,
  Eur.\ Phys.\ J.\ C {\bf 71} (2011) 1515
   [Erratum-ibid.\ C {\bf 72} (2012) 1874]
  [arXiv:1010.4180 [hep-ph]].
%
\bibitem{HMNT}
  K.~Hagiwara, R.~Liao, A.~D.~Martin, D.~Nomura and T.~Teubner,
  J.\ Phys.\ G G {\bf 38} (2011) 085003
  [arXiv:1105.3149 [hep-ph]].
%
\bibitem{Benayoun:2012wc}
  M.~Benayoun, P.~David, L.~DelBuono and F.~Jegerlehner,
  arXiv:1210.7184 [hep-ph].
%
\bibitem{JegerlehnerSzafron}
  F.~Jegerlehner and R.~Szafron,
  Eur.\ Phys.\ J.\ C {\bf 71} (2011) 1632
  [arXiv:1101.2872 [hep-ph]].
%
\bibitem{dRPV}
  J.~Prades, E.~de Rafael and A.~Vainshtein,
  arXiv:0901.0306 [hep-ph].
%
\bibitem{Goecke:2010if}
  T.~Goecke, C.~S.~Fischer and R.~Williams,
  Phys.\ Rev.\ D {\bf 83} (2011) 094006
   [Erratum-ibid.\ D {\bf 86} (2012) 099901]
  [arXiv:1012.3886 [hep-ph]].
%
\bibitem{Blum:2013qu}
  T.~Blum, M.~Hayakawa and T.~Izubuchi,
  arXiv:1301.2607[hep-ph].
%
%
\bibitem{CzMV}
  A. Czarnecki, W.~J. Marciano, A. Vainshtein
  Phys.Rev.D {\bf 67} (2003) 073006, Erratum-ibid.D{\bf 73} (2006)
  119901.
%
\bibitem{ATLAS:2013mma}
  [ATLAS Collaboration],
  ATLAS-CONF-2013-014.
%
\bibitem{CMS:yva}
  [CMS Collaboration],
  CMS-PAS-HIG-13-005.
%
\bibitem{PDG2012}
  J.~Beringer et al. (Particle Data Group)
  Phys.\ Rev.\ D {\bf 86} (2012) 010001.
\bibitem{Awramik:2003rn}
  M.~Awramik, M.~Czakon, A.~Freitas and G.~Weiglein
  Phys.\ Rev.\ D {\bf 69} (2004) 053006
\bibitem{Ferroglia:2012ir}
  A.~Ferroglia and A.~Sirlin,
  Phys.\ Rev.\ D {\bf 87} (2013) 037501. 
\bibitem{CKM2}
  A.~Czarnecki, B.~Krause and W.~J.~Marciano,
  Phys.\ Rev.\ Lett.\  {\bf 76} (1996) 3267
  [hep-ph/9512369].
\bibitem{HSW04}
  S.~Heinemeyer, D.~St\"ockinger and G.~Weiglein,
  Nucl.\ Phys.\ B {\bf 699} (2004) 103.
%
\bibitem{CzarneckiGribouk}
  T.~Gribouk and A.~Czarnecki,
  Phys.\ Rev.\ D {\bf 72} (2005) 053016
  [hep-ph/0509205].
%
\bibitem{CKM1}
  A.~Czarnecki, B.~Krause and W.~J.~Marciano,
  Phys.\ Rev.\ D {\bf 52} (1995) 2619
%
\bibitem{Cheung:2001hz}
  K.~-m.~Cheung, C.~-H.~Chou and O.~C.~W.~Kong,
  Phys.\ Rev.\ D {\bf 64} (2001) 111301
  [hep-ph/0103183].
\bibitem{Wu:2001vq}
  Y.~-L.~Wu and Y.~-F.~Zhou,
  Phys.\ Rev.\ D {\bf 64} (2001) 115018
  [hep-ph/0104056].
%
\bibitem{HSW03}
  S.~Heinemeyer, D.~St\"ockinger and G.~Weiglein,
  Nucl.\ Phys.\ B {\bf 690} (2004) 62.
%
\bibitem{BarrZee}
  S.~M.~Barr and A.~Zee,
  ``Electric Dipole Moment Of The Electron And Of The Neutron,''
  Phys.\ Rev.\ Lett.\  {\bf 65} (1990) 21
  [Erratum-ibid.\  {\bf 65} (1990) 2920].
%
\bibitem{Stockinger:2006zn}
  D.~St\"ockinger,
  ``The Muon Magnetic Moment and Supersymmetry,''
  J.\ Phys.\ G{\bf 34} (2006) R45-R92.
%
\bibitem{Knechtreview}
  M.~Knecht,
  Lect. \ Notes \ Phys. {\bf 629} (2004) 37-84
%
\bibitem{DG98}
  G.~Degrassi and G.~F.~Giudice,
  Phys.\ Rev.\ D {\bf 58} (1998) 053007.
%
\bibitem{Peris:1995bb}
  S.~Peris, M.~Perrottet and E.~de Rafael,
  Phys.\ Lett.\ B {\bf 355} (1995) 523
  [hep-ph/9505405].
\bibitem{Knecht:2002hr}
  M.~Knecht, S.~Peris, M.~Perrottet and E.~De Rafael,
  JHEP {\bf 0211} (2002) 003
  [hep-ph/0205102].

\end{thebibliography}
\end{document}